# Intelligent Cause Prioritisation?

# An Analysis of AI Policy Priorities and Governance in Africa

Osaremen Iluobe
oiluobe@sas.upenn.edu

Kisso Selvan
ks975@cantab.ac.uk

**Abstract**

The rapid improvement of AI systems has intensified debate about humanity's economic, political, social, and existential future. As AI reshapes expectations about what lies ahead, policy choices and institutional responses will play a crucial role in determining who benefits, who bears the costs, and whether the most serious risks can be mitigated. Africa remains relatively overlooked in these discussions, partly because it is largely a consumer rather than a producer of frontier AI systems, and also because many countries on the continent continue to face pressing development challenges. Given the catch-up imperative, governments across Africa are eager to embrace AI as a means for accelerating economic transformation. Drawing on speeches, press releases, public statements, and national AI strategies/frameworks, this essay argues that while African governments are highly attentive to AI's opportunities, they devote comparatively little attention to AI safety.

*We've benefited from helpful comments given to us by Dr. Chinasa T. Okolo, Dr. Nikita Makarchev, and Professor Jonathan Shock. We'd also like to extend our thanks to attendees at the 2026 Effective Altruism Global Conference in London for inspiring further thoughts on this topic.

***"Our Slumber Worries Me More Than AI"***
***— Abiy Ahmed (2026), Prime Minister of Ethiopia[1]***

***"A more moral AI is not enough if that morality is determined by a few. What is needed is a more active political involvement that is capable of slowing things down when everything is accelerating…"***
***— Pope Leo XIV (2026)[2]***

---

## Introduction

Artificial intelligence (AI) is reshaping the conditions under which economic development, political competition, and social organisation take place. The pace of this shift has sharpened debates over distribution, including who benefits, who bears the costs, and whether the most severe risks can be managed before they become entrenched. Given that the institutions, regulations, and governance frameworks being established today will shape AI trajectories for decades, this matters a great deal. Countries that engage early in setting these rules are more likely to influence outcomes than those that remain peripheral.

Africa occupies an uneasy position within this landscape. The continent remains largely a consumer rather than a producer of frontier AI systems (Pilling, 2024; Meka, 2025) and many states continue to face pressing development challenges. At the same time, AI is arriving regardless through national strategies, foreign investment, and the rapid diffusion of general-purpose tools, and African governments are actively responding. Across the continent, policymakers have embraced AI as a pathway to economic transformation, infrastructure leapfrogging, and improved public service delivery (Malatji, 2025; Aryee et al., 2025). This optimism is genuine and, in many respects, justified.

But what is less clear is whether this enthusiasm is matched by sufficient attention to the risks AI introduces. A growing literature documents trends in Africa's AI policy landscape (Kwarkye, 2025) as well as near-term harms already emerging in African contexts, including electoral disinformation, algorithmic bias, surveillance misuse, and labour displacement (Segun et al., 2025; Gwagwa et al., 2020). Less visible in policy discourse are longer-horizon structural risks, such as the concentration of AI power, the erosion of democratic accountability, and increasing dependence on physical and institutional infrastructure designed elsewhere (Weitzel, 2026; Khanal et al., 2025).

This essay examines this gap semi-empirically. Drawing on a corpus of 35 speeches, press releases, and public statements issued between 2023 and 2026, supplemented by a qualitative assessment of national AI strategies from twelve countries and the African Union, it analyses how African governments view AI's opportunities and risks. Developmental framing is universal and dominant, while safety considerations are uneven, comparatively shallow, and focused overwhelmingly on near-term concerns. The institutional mechanisms, monitoring systems, and international coordination commitments required to address longer-term risks are largely absent. Where safety language does appear in strategy documents, it likely exceeds the underlying institutional capacity to implement it (Ireri et al., 2026; Diallo et al., 2025). It should be stressed that while Africa is not unique in prioritising adoption over governance, the combination of structural vulnerability, constrained regulatory capacity, and limited representation in global AI governance forums makes the safety-adoption gap especially consequential.

The essay proceeds as follows. Section 2 outlines the conceptual framework, distinguishing between proximate and structural AI safety concerns and situating Africa's position within the broader literature on AI governance. Section 3 describes the data and methods. Section 4 presents the empirical findings, organised around two patterns in the corpus. Section 5 discusses implications, with attention to lock-in effects, path dependency, and the collective security dimensions of AI risk. Section 6 concludes.

## 2. Conceptual Framework

How African governments understand the risks of advanced AI, and how they ought to approach AI safety, forms the focus of this section and recurs throughout the essay.

### 2.1 Proximate and Structural AI Safety

AI safety is neither a singular nor a simple concept. Rather, it encompasses a range of concerns that differ in time horizon, tractability, and the institutional responses they require. Given this and for the purposes of this essay, a heuristic demarcation is drawn between 'proximate' and 'structural' AI safety. Proximate safety refers to near-term, relatively tractable risks arising from the deployment of existing AI systems. These include the use of AI to generate and amplify disinformation, manipulate electoral processes, enable surveillance, perpetuate algorithmic bias, and facilitate cybercrime. These risks are already materialising. For instance, during Kenya's 2022 elections, altered audio clips and social media bots contributed to the amplification of ethnic tensions. OpenAI's 2025 threat reporting also documented attempts to use generative AI to influence Ghana's presidential election (Segun et al., 2025). In response to concerns which lie at the intersection of AI, misinformation and political processes, several African governments have taken measures or issued statements of intent to tackle the growing problem. Proximate safety concerns can, in principle, be addressed through domestic regulation, platform governance, and firm cooperation. They are also more politically visible, as the harms are easier to observe and often affect politicians directly through deepfakes and AI-generated misinformation.

Structural safety refers to longer-term and more structurally difficult risks that may arise as AI systems become more capable. These include the concentration of AI-enabled power in a small number of states or corporations, the weakening of democratic accountability and human oversight, and the possibility that advanced systems pursue goals misaligned with human welfare (Bengio et al., 2024; Carlsmith, 2024). Davidson et al. (2025) argue that sufficiently advanced AI could introduce new forms of power seizure without historical precedent, while Khanal et al. (2025) show that frontier AI development is already becoming concentrated in a small set of institutions whose influence increasingly resembles that of sovereign actors. These risks extend across borders, cannot be handled through domestic regulation alone, and are difficult to address after they emerge. Hence, they require international coordination, stronger governance institutions at the frontier of AI development, and meaningful participation in global rule-making.

This dichotomy is not, to our knowledge, an established distinction in the AI safety literature, but it builds directly on existing taxonomies. The structural category follows Zwetsloot and Dafoe (2019), who separate risks arising from how AI reshapes incentives, power dynamics, and institutional environments from those caused by accidents or misuse, while the pairing parallels the near-term/long-term distinction in AI ethics, mindful of Prunkl and Whittlestone's (2020) caution that such labels can conflate capability, immediacy, certainty, and severity. Both categories also map onto the MIT AI Risk Repository, a meta-review of over 1,700 risks drawn from 74 existing frameworks (Slattery et al., 2024): proximate safety corresponds broadly to its misinformation, discrimination and toxicity, and privacy and security domains, while structural safety corresponds to its socioeconomic and environmental domain, most notably the subdomains of power centralisation and governance failure. Two clarifications are warranted. First, the labels are descriptive rather than hierarchical: neither category is prior to, or more important than, the other. Second, they should not be confused with soft law and hard law, which in AI governance denote non-binding and binding regulatory instruments respectively.

Although a distinction is made here, structural and proximate AI safety are not entirely separate and are certainly not unrelated as weaknesses in proximate safety can increase structural safety risks. For instance, widespread disinformation can undermine democratic accountability, weaken oversight, and make power concentration more likely (Bengio, 2023). It is therefore important to treat proximate and structural safety as complementary rather than competing priorities. At the same time, the distinction matters in the African context because each places different demands on institutions and international engagement. Proximate safety can be partially managed through existing regulatory tools, but structural safety cannot be effectively governed from national parliaments or presidential palaces alone.

### 2.2 The Catch-Up Imperative and the Safety Deficit

African governments enter the AI era from a position of relative material and institutional disadvantage. It is for this reason that some have argued that “Africa should not even talk about AI when access to common electricity is not available on the continent" (Elumelu, 2025). Despite some scepticism, the idea of leapfrogging, where AI enables countries to bypass traditional development pathways in a manner similar to mobile phones, has been met with enthusiasm by policymakers and has some empirical support in sectors such as healthcare, agriculture, and financial services (Pilling, 2024).

While the leapfrogging narrative is central to the push for AI adoption for many African governments, it can also create blind spots around more serious safety concerns. Indeed, one way to interpret the resulting gap in attention is that it reflects rational prioritisation under severe resource constraints. In this respect, AI safety is considered something of a luxury good and only becomes a priority once basic development needs are met. Under this interpretation, governments facing urgent shortages in health, education, and infrastructure naturally direct limited capacity away from risks that appear uncertain or distant. While this explanation has some merit, it is incomplete for two reasons.

First, the most important AI risks for African countries are not temporally distant. Disinformation, surveillance misuse, and labour displacement are already present, while longer-term risks linked to technological dependence and governance lock-in are building up now (Ireri et al., 2026; Gwagwa et al., 2020). This weakens the analogy to deferred luxury consumption. Second, the idea of a trade-off between safety and opportunity is not fixed, but shaped by political and institutional context. AI discourse in Africa is often framed in strongly optimistic terms that downplay constraints, while Khanal et al. (2025) show that large technology firms actively shape policy agendas by influencing which issues are treated as priorities. In this sense, the marginalisation of safety is partly produced by external influence and by incentives that favour adoption.

## 3. Data and Method

The empirical analysis draws on two complementary sources. The first is a corpus of 35 speeches, press releases, and public statements issued by African governments and the African Union between 2023 and 2026. The corpus covers ten states, Nigeria, South Africa, Kenya, Ghana, Rwanda, Ethiopia, Zambia, Côte d'Ivoire, Morocco, and Uganda, alongside the African Union. It spans the period in which AI discourse became more prominent on a global level following the launch of ChatGPT in November 2022. The second source is a set of national AI strategies and policy frameworks from thirteen countries, Nigeria, Ethiopia, Ghana, Côte d'Ivoire, Senegal, South Africa, Kenya, Rwanda, Egypt, Zambia, Benin, Morocco, and Mauritania, alongside the African Union, based on documents available publicly as of mid-2026.

It should be noted that the national strategies/policy frameworks often exhibit an abundance bias, as these documents often contain aspirational or visionary language that may exceed near-term implementation capacity. The empirical analysis is therefore deliberately parsimonious and based on how governments publicly present AI in official documents and communications, rather than a direct measure of implementation or governance capacity. This matters because public framing can shape agendas and define what counts as a policy problem, but it does not necessarily reflect what is implemented in practice. The speeches, statements, and press releases corpus is also not fully representative as it is restricted to occasions where AI is a central focus which is contingent on occasion and audience, and further limited by the exclusion of some materials due to language and translation constraints. Overall, what the respective corpora reflect are the relative priorities of governments across Africa as they relate to AI but the snapshot provided by that which is analysed is an effective way of gauging the AI policy landscape as it unfolds.

A further clarification concerns the choice of method. Computational text-as-data approaches, such as topic modelling or supervised classification, were considered but set aside. This owes to the fact that statements in which AI is a central focus remain scarce among African governments, and the resulting corpus is too small and too heterogeneous in genre, length, and language to support stable topic estimation or classifier training. A quasi-empirical analysis was nonetheless retained. Each document was instead read in full and coded qualitatively against the explicit benchmarks described below, which preserves systematic, corpus-wide comparability while remaining appropriate to a small-N corpus.

National strategy and/or policy frameworks were coded along four dimensions. Tone captures how AI is framed overall, distinguishing between Enthusiastic (AI as opportunity), Instrumental (AI as a development tool requiring management), and Cautious (AI as a domain requiring restraint). Development Emphasis measures the extent to which documents present AI as a driver of economic transformation and public sector modernisation. Safety and Risk captures the prominence of governance and risk considerations, distinguishing between proximate safety, structural safety, and minimal engagement with either. Framework Depth, applied to strategy documents, assesses institutional specificity, including implementation mechanisms, monitoring systems, and enforcement tools, as distinct from high-level or aspirational language.

Similarly, speeches, statements and press releases made by state leaders and ministers are also grouped into three inductively derived criteria. Development Champions frame AI primarily as an engine of growth and digital transformation, with limited attention to risk. Governance Advocates place greater emphasis on ethics, regulation, and responsible deployment. Strategic Balancers combine strong developmental ambition with more substantive engagement on governance concerns. These postures are summarised in Tables 1 and 2 and form the basis for the analysis provided in Section 4.

## 4. Findings

The corpus reveals two consistent patterns. First, developmental framing is universal across the sample. Second, engagement with AI safety is uneven and focused primarily on near-term risks.

### 4.1 Development Emphasis is Universal

Across all countries examined, AI is presented as a tool for structural transformation, public-sector modernisation, and participation in the global digital economy. This is perhaps unsurprising given the potentially transformative capabilities of AI. Furthermore, AI-optimism more generally is not a uniquely African phenomenon.

As for the specific priorities and ambitions of countries, Nigeria links AI directly to its ambition of becoming a trillion-dollar economy, Rwanda positions itself as a continental AI hub, Ethiopia presents AI as a pathway to becoming a centre of excellence in Africa, and Senegal integrates AI into its broader national development agenda. Similar patterns are visible in countries with more limited digital capacity, including Benin and Mauritania. Overall though, despite differences in income, political systems, and state capacity, development emphasis remains consistently high. The uniformity of this discourse suggests that the catch-up imperative functions as a powerful organising principle in African AI governance. AI is understood first as a development opportunity, with questions of risk generally taking a subordinate position.

Table 1: Comparative Characteristics of African National AI Strategies

| Country | Tone | Development Emphasis | Safety | Framework Depth |
|---|---|---|---|---|
| Nigeria | Instrumental | High | High | Substantive |
| Ethiopia | Instrumental | High | Med–High | Substantive |
| Ghana | Enthusiastic | High | Med–High | Substantive |
| Côte d'Ivoire | Enthusiastic | High | Low | Substantive |
| Senegal | Enthusiastic | High | Low | Substantive |
| South Africa | Cautious | High | Medium | Substantive |
| Kenya | Instrumental | High | Med–High | Substantive |
| Rwanda | Instrumental | High | Med–High | Substantive |
| Egypt | Enthusiastic | High | Low | Substantive |
| Zambia | Instrumental | High | Low | Substantive |
| Benin | Instrumental | High | Low | Substantive |
| Morocco | Cautious | High | High | Substantive |
| Mauritania | Instrumental | High | Low | Moderate |
| African Union | Instrumental | High | Med-High | Substantive |

*Notes:* Tone captures the overall framing of AI (Enthusiastic, Instrumental, or Cautious). Development Emphasis reflects the extent to which AI is presented as a tool for economic transformation and public-sector development. Safety captures the prominence of ethical, governance, and risk-management concerns. Framework Depth assesses the level of institutional specificity, implementation mechanisms, and monitoring provisions contained in the strategy.

### 4.2 Safety Engagement is Uneven and Concentrated on Proximate Risks

In contrast to the near-universal emphasis on development, engagement with safety concerns varies considerably across countries. South Africa, Kenya, and Morocco place relatively greater emphasis on governance and responsible deployment, while Ethiopia, Zambia, Côte d'Ivoire, and Senegal focus much more heavily on adoption and economic opportunity. Morocco has gone further than any other African country in emphasising the risks associated with advanced AI, both in its policy framework and public statements, as shown in Table 2. Nigeria, Ghana, and Rwanda occupy an intermediate position in that they combine ambitious development goals with more substantive discussion of governance issues.

Across national frameworks and public statements/press releases, where safety concerns appear, they are overwhelmingly concentrated on proximate safety risks, particularly disinformation, data privacy, algorithmic bias, cybersecurity, and labour-market disruption. Kenya and Nigeria stand out in this regard. Kenya's strategy identifies data colonialism, labour displacement, and platform power as governance challenges, while Nigeria's strategy includes a formal risk assessment framework and explicit risk categorisation. As such, what is largely absent across the corpus is sustained engagement with structural safety concerns. Issues such as power concentration, democratic erosion, technological dependency, and risks associated with increasingly capable frontier AI systems receive little attention in either strategy documents or public communications. Although the AU Continental AI Strategy acknowledges the need for governance and regulation, structural safety concerns remain peripheral to the broader policy conversation as evidenced by the public statements and national policy frameworks which African governments have issued.

Table 2: Classification of AI Speech and Press Release Narratives

| Country | Tone | Development | Safety | Speech Posture |
|---|---|---|---|---|
| Nigeria | Enthusiastic / Instrumental | High | Medium | Strategic Balancer |
| South Africa | Mixed | High | Medium–High | Governance Advocate |
| Kenya | Instrumental | High | Medium–High | Governance Advocate |
| Ghana | Enthusiastic | High | Medium | Strategic Balancer |
| Rwanda | Instrumental | High | Medium | Strategic Balancer |
| Ethiopia | Enthusiastic | High | Low | Development Champion |
| Zambia | Enthusiastic / Instrumental | High | Low | Development Champion |
| Côte d'Ivoire | Enthusiastic | High | Low | Development Champion |
| Morocco | Cautious | Medium | High | Governance Advocate |
| Uganda | Enthusiastic | High | Low | Development Champion |
| African Union | Instrumental | High | Medium–High | Governance Advocate |

*Notes*: Development Champions frame AI primarily as an opportunity for growth, investment, and digital transformation, with limited attention to risks. Governance Advocates place greater emphasis on ethics, regulation, inclusion, and responsible deployment. Strategic Balancers combine strong developmental ambitions with substantive engagement with governance and risk concerns.

## 5. Discussion

The findings raise two broader questions. Why do African governments consistently largely overlook AI safety? And what are the consequences of this pattern for Africa's position in the emerging AI order?

### 5.1 Explaining the Pattern

A natural explanation is resource constraint. Governments facing urgent challenges in health, education, infrastructure, and energy may reasonably view AI safety as a secondary concern that can wait until more immediate development priorities are addressed. As a consequence and given that AI is widely believed to engender significant material gains, a fixation on how those potential gains can be realised and harnessed is unsurprisingly the main priority of governments across Africa. However, this may prove negatively consequential as once AI is established primarily as a development opportunity, concerns about robust governance and safety can easily appear secondary or even obstructive. As Meka (2025) argues, African AI discourse is characterised by a persistent optimism that often understates structural constraints. The corpus suggests that this optimism shapes not only what governments prioritise, but also what they perceive as requiring governance in the first place.

There is also a more practical reason why structural safety receives less attention. Near-term risks such as disinformation, algorithmic bias, and data privacy violations fit within familiar regulatory frameworks. They have identifiable victims and generate domestic political pressure. Structural safety concerns are different. Questions of power concentration, technological dependency, or AI-enabled political control are more diffuse, more uncertain, and harder to translate into concrete policy interventions. Even governments that recognise these risks may struggle to convert concern into action. This does not justify inaction, but it helps explain why the gap between optimism and safety, as well as between proximate and structural safety engagement, is so pronounced.

### 5.2 Why It Matters

The implications of this pattern extend beyond current policy debates. The first concerns lock-in. Technologies rarely arrive on a blank institutional slate. Once deployment patterns, infrastructure investments, and governance arrangements become established, they are difficult to reverse. Governments that postpone engagement with longer-term AI risks may therefore be narrowing the range of options available to them in the future. Decisions made today about infrastructure, standards, and governance relationships will likely shape the development trajectory of AI for years to come.

The second implication relates to Africa's position in global AI governance. Structural safety is not a problem that any country can address alone. Risks associated with advanced AI, including power concentration, loss of oversight, and technological dependency, require international coordination (Bengio et al., 2024; Davidson et al., 2025). Yet the institutions and forums shaping those discussions remain dominated by a small number of actors, primarily the United States, the United Kingdom, the European Union, and a handful of Asian economies. Africa's participation remains limited. The continent accounts for less than 0.5 percent of global machine learning and large language model production (Pilling, 2024), and only one African country currently participates in the AI Safety Institute International Network (Ireri et al., 2026). In important respects, the governance architecture for advanced AI is being built without substantial African input.

This matters because governance frameworks are not neutral. Standards developed in high-resource environments are unlikely to fully reflect the realities of African deployment contexts. Ireri et al. (2026) show that safety evaluations designed elsewhere do not transfer reliably to African settings, where infrastructure constraints, state capacity limitations, and linguistic diversity create distinct challenges. The risk is therefore not only technological dependency, but governance dependency as well. Rules developed elsewhere may come to shape African AI systems without adequately reflecting African priorities or vulnerabilities.

As these risks become clearer, the imbalance between development and safety becomes increasingly difficult to defend. AI systems that are poorly governed are unlikely to distribute benefits broadly. Rather they risk concentrating gains among those already best positioned to capture them while shifting costs onto those least able to bear them (Weitzel, 2026; Khanal et al., 2025). If African governments wish to shape the future of AI rather than simply adapt to it, engagement with AI safety will need to be treated not as a concession to external priorities, but as part of the broader project of ensuring that AI develops on terms consistent with African interests and realities.

## 6. Conclusion

It should be highlighted that Africa is not absent from the global AI conversation. Across the continent, governments are producing strategy documents, convening stakeholder consultations, and issuing public commitments to AI-driven development which indicates a seriousness towards the topic. Indeed, the countries examined in this essay are, by continental standards, already ahead given that the overwhelming majority of African states have no national AI strategy or governing framework whatsoever. The findings should be read in that light. As such, this essay is not a verdict on African AI governance broadly. More accurately, it is an assessment of the continent's most engaged actors, and if even they are underweighting safety, the picture across the continent as a whole is likely worse.

What this essay fundamentally argues is that there is currently a tendency to lead with opportunity and relegate risk. If the perceived trade-off between innovation and safety contributes to the relative neglect of AI safety, we argue that African governments should recognise that this trade-off is largely artificial. Additionally, where risks are addressed in strategy documents and via public communication, they are primarily focused on disinformation, algorithmic bias, and data privacy. Whilst these are genuine problems and it matters that African governments are beginning to address them, considerations of the structural risks associated with advanced AI are largely absent from the policy conversation.

The implications from the findings of this paper are twofold. First, there should be greater consideration given to safety as a whole by governments across Africa, especially structural risks which are seemingly given no attention at all in existing policy and strategy frameworks. Second, there should be an urgency to move away from the current framing which is all-consumed by a catch-up imperative. The institutions, standards, and coordination mechanisms governing advanced AI are being established now and by a small number of actors. Lock-in effects mean that African governments cannot afford to be sidelined and should therefore engage proactively in global AI safety forums to ensure that emerging safeguards also reflect their concerns. African governments have consistently insisted they will not be passive actors in the AI revolution. The evidence in this essay suggests that posture has not yet extended to AI safety. It needs to.

## Notes

1. Ethiopian Broadcasting Corporation (2026) *'Our Slumber Worries Me More Than AI,' Says Ethiopian Prime Minister*. Available at: https://www.ebc.et/english/Home/NewsDetails?NewsId=3840 (Accessed: 13 March 2026).

2. Holy See (2026) *Encyclical Letter: Magnifica Humanitas of His Holiness Pope Leo XIV on Safeguarding the Human Person in the Time of Artificial Intelligence*. Available at: https://www.vatican.va/content/leo-xiv/en/encyclicals/documents/20260515-magnifica-humanitas.html (Accessed: 2 June 2026).